\documentclass[preprint,showpacs,pre,prb,amsmath,amssymb,endfloats*]{revtex4}
\usepackage{graphicx}
\usepackage{dcolumn}
\begin{document}

\title{Calorimetric study of the nematic to smectic-A phase transition in octylcyanobiphenyl-hexane binary mixtures }

\author{Krishna P. Sigdel and Germano S. Iannacchione\footnote{electronic address: gsiannac@wpi.edu}}
\affiliation{Department of Physics, Worcester Polytechnic
Institute, Worcester, Massachusetts 01609, USA}

\date{\today}


\begin{abstract}
The continuous nematic to smectic-$A$ ($N$-Sm$A$) phase transition
has been studied by high-resolution ac-calorimetry in binary
mixtures of the liquid crystal octylcyanobiphenyl(8CB) and a
non-mesogenic, low-molecular weight, solvent n-hexane(hex) as a
function of temperature and solvent concentration. Heating and
cooling scans about the $N$-Sm$A$ transition temperature were
repeatedly performed on pure and six 8CB+hex samples having hexane
molar concentration ranging from $x_{hex}= 0.02$ to $0.12$. All
8CB+hex samples in this range of $x_{hex}$ remain macroscopically
miscible and exhibit an $N$-Sm$A$ heat capacity peak that shifts
non-monotonically to lower temperature and evolves in shape, with
a reproducible hysteresis, as $x_{hex}$ increases. The imaginary
part of heat capacity remains zero up to $x^{TCP}_{hex}\simeq
0.07$ above which the distinct peak is observed, corresponding to
a jump in both the real and imaginary enthalpy. A simple power-law
analysis reveals an effective exponent that increases smoothly
from $0.30$ to $0.50$ with an amplitude ratio
$A^{-}/A^{+}\rightarrow 1$ as $x_{hex}\rightarrow x^{TCP}_{hex}$.
This observed crossover towards the $N$-Sm$A$ tricritical point
driven by solvent concentration is consistent with previous
results and can be understood as weakening of the liquid crystal
intermolecular potential promoting increased nematic fluctuations.
\end{abstract}

\pacs{61.30.-v, 64.70.mj, 65.40.Ba}

\maketitle


\section{INTRODUCTION}\normalfont

\label{sec:intro} Liquid crystals(LCs) are anisotropic fluids
which exhibit a varieties of phases and phase transitions
\cite{deGennes93, Chandrashekhar92}. The nematic and smectic-$A$
phases are the best known phases of liquid crystals. The
transition between nematic ($N$) and smectic-$A$ (Sm$A$) phases is
interesting and important because it involves the breaking of a
continuous symmetry as well as sharing some properties with the
superconducting transition in metals and the superfluid transition
in $^{4}He$. The $N$-Sm$A$ transition is also a model phase
transition for the study of confinement and disorder effects such
as in mixtures with silica aerosil
\cite{Germano03,Germano04,Floren09},
 or embedded in an aerogel \cite{Wu95, Bellini01} and controlled porous
 glass \cite{Zidan04, Kutnjak03, Kutnjak04}. The phase transition behavior is also
 sensitive to an applied external electric and  magnetic field \cite{Ranjan99,Andrew02}
 as well as with LC+LC mixtures \cite{Stine89,Sied02,Lafouresse03,Sied,Jeong08}.
 Even though the $N$-Sm$A$ transition has been extensively studied
\cite{Garland94}, there remains many unresolved issues regarding
the fundamental nature of the transition.

Recently, attention has been drawn to the study of miscible
mixtures of liquid crystals and non-mesogenic, low-molecular
weight, solvents for broadening the basic understanding of
mesogenic order, critical behavior and tuning viscoelastic
properties
\cite{Denolf07,Denolf06,Mukherjee02,DasGupta01,Rieker95}. X-ray
diffraction experiments performed on smectic-$A$ and smectic-$C$
thermotropic liquid crystals have demonstrated that the smectic
layer spacing increases with the addition of organic solvents to
the host liquid crystal indicating the formation of an organic
lyotropic lamellar liquid crystal phase\cite{Rieker95}. It was
suggested from the visual inspection that for octylcyanobiphenyl
(8CB) and n-hexane(hex) mixture systems ( having a volume fraction
of $\geq 0.1$ ), the solvent is not uniformly distributed
throughout the host LC and minimal, non-reproducible, swelling
occurs. It was also suggested that the amount of solvent
incorporated in a smectic liquid crystal depends on the host
liquid crystal, nature and amount of solvent, and temperature;
noting that the mixture phase separates for a solvent to liquid
crystal mole ratio $\geq 1.0$. Other studies of the effect of a
biphenyl solvent on the splay and bend elastic constant and the
rotational viscosity coefficient of 8CB observed an anomalous
behavior of $K_{11}$ and $\Delta \varepsilon$ near $N$-Sm$A$
transition\cite{DasGupta1}. A theoretical study on the influence
of non-mesogenic solvent on the $N$-Sm$A$ phase transition using
Landau approach found a concentration induced tricritical point
for the $N$-Sm$A$ transition\cite{Mukherjee02}. This theoretical
model also found that the Frank elastic constants $K_{11}$,
$K_{22}$, and $K_{33}$ are modified as a function of solvent
concentration near the $N$-Sm$A$ phase transition.

A recent calorimetric study of the $N$-Sm$A$ transition in
mixtures of 8CB and cyclo-hexane (8CB+chex) was performed under
continuous stirring conditions \cite{Denolf07}. This study found a
linear decrease of the transition temperatures $T_{NA}$ with a
linear increase of critical heat capacity exponent $\alpha$ with
increasing mole fraction of cyclohexane $x_{chex}$. This behavior
ends at a tricritical point (TCP) where the transition becomes
first-order at $x^{TCP}_{chex}=0.046$, just below which $\alpha =
0.5$ and the nematic range $\Delta T_{N} = T_{IN}-T_{NA}= 4.8K$.
For $x_{chex} > x^{TCP}_{chex}$, the $N$-Sm$A$ latent heat
smoothly increases non-linearly from zero \cite{Denolf07}.

The $N$-Sm$A$ phase transition is a non-trivial member of the
3D-XY universality class due to the anisotropy of its critical
fluctuations parallel and perpendicular to the director
\cite{Garland94, Yethiraj02, Garland08}. The $N$-Sm$A$ critical
behavior is strongly effected by the coupling between the smectic
order parameter $\psi(\vec{r})=\psi_{0}\exp(i\vec{q_{0}}.\vec{r})$
and the nematic
 order parameter $Q_{ij}=(1/2)S\left(3\hat{n}_{i}\hat{n_{j}}-\delta_{ij}\right)$.
Here, the $\psi$ is the amplitude of the one-dimensional density
wave, $\rho(\vec{r})=
Re[\rho_{0}+\exp(i\vec{q_{0}}.\vec{r})\psi(\vec{r})]$,
$q_{0}=2\pi/d$ is the wave vector corresponding to the layer
spacing $d$, $S$ is a scalar parameter measuring the magnitude of
orientational order on short length scales, and $\hat{n}$ is the
nematic director describing spatial orientation of the
orientational axis on longer length scales. It has been shown by
de Gennes \cite{deGennes93} and McMillan \cite{McMillan71} that a
mean-field coupling between $S$ fluctuations and smectic order
$\delta S-\psi$ can drive a second-order $N$-Sm$A$ phase
transition first-order via a tricritical point. The theory
proposed by Halperin, Lubenski, and Ma (HLM)
\cite{Halperin74,Anisimov90}, taking into account the coupling
between $\psi$ and the nematic director fluctuations
$\delta\hat{n}$, showed that the $N$-Sm$A$ transition is always at
least weakly first order which rules out the possibility of a
tricritical point. Combining both the $\delta S$ and
$\delta\hat{n}$ couplings introduces two more terms in
 free energy  expression as compared to the usual standard form.
One term is of the form $\psi^{2}S$ which is nematic-smectic order
parameter coupling (referred to as de Gennes coupling) and the
other is smectic order-nematic director fluctuation coupling (HLM
coupling) $\psi^{2}\delta\hat{n}$. The former coupling reveals the
effects of the elasticity of the nematic order prior to the onset
of the smectic order and can drive the $N$-Sm$A$ transition from
XY like to tricritical to weakly first-order\cite{Germano03}. The
coupling $\psi^{2}\delta\hat{n}$ causes the anisotropic elastic
deformations in the smectic. The strength of this coupling depends
on the magnitude of the splay elastic constant $K_{11}$
 which is directly proportional to $S^{2}$. Since it is expected that a low-molecular weight solvent miscible in an LC would affect
 both $\delta S$ and $\delta\hat{n}$ fluctuations, the $x_{sol}$
 dependence would be accounted for using similar terms in a
 free-energy expansion.

In this work, the effect of a non-mesogenic, low molecular weight,
solvent (n-hexane) concentration on the continuous nematic to
smectic-$A$ ($N$-Sm$A$) phase transition on octylcyanobiphenyl
(8CB) and n-hexane (hex) binary mixtures (8CB+hex) was studied via
high-resolution ac-calorimetry as a function of n-hexane
concentration, $x_{hex}$. The introduction of n-hexane on 8CB
causes a dramatic change in the $N$-Sm$A$ phase transition
behavior. The heat capacity peaks associated with the $N$-Sm$A$
transition, $\delta C_{p}$, shift towards lower temperature
non-monotonically and become progressively larger as the hexane
concentration increases. The dispersive part of heat capacity
$C^{"}_{p}$ associated with $N$-Sm$A$ transition has peaks only
for higher hexane mole fractions $(x_{hex}\geq 0.08)$ but not for
the lower hexane mole fractions $(x_{hex}\leq 0.06)$ revealing the
continuous (second order) nature of the $N$-Sm$A$ transition for
the lower n-hexane mole fractions $(x_{hex}\leq 0.06)$ and
first-order nature for higher n-hexane mole fractions
$(x_{hex}\geq 0.08)$. The integrated ac-enthalpy increases overall
as a function of hexane molar fraction whereas the imaginary part
of the enthalpy reveals a sharp increase at hexane mole fraction
of around $0.07$ and remains fairly constant. The crossover
between continuous to first-order $N$-Sm$A$ transition is observed
at a tricritical point of $x^{TCP}_{hex}\approx 0.07$. The
non-linear increase in the heat capacity effective critical
exponent towards its tricritical value $(\alpha=0.5)$ is observed.

The hysteresis of the $\delta C_{p}$ shape on heating and cooling
has been observed is likely due to a microscopic phase separation
of the solvent, perhaps into intersticial region between smectic
layers. The non-monotonic transition temperature shift may be due
to the competing interactions of microphase separation and
dilution effects. These effects may also responsible for the
$\alpha_{eff}$ behavior with extended curvature as
$x_{hex}\rightarrow x^{TCP}_{hex}$. These effects would also have
profound consequences on the higher temperature $I$-$N$ phase
transition as well, which was presented in previous paper
\cite{krishna1}.

 This paper is organized as follows; following this introduction,
Section~\ref{sec:exp} describes the preparation of sample, the
calorimetric cell, and the ac-calorimetric procedures employed in
this work. Section~\ref{sec:results} describes the calorimetric
results and critical behavior of the $N$-$SmA$ phase transition in
the 8CB+hex system. Section~\ref{sec:disc} discusses these results
and draws conclusions.


\section{Experimental}
\label{sec:exp}

The liquid crystal 8CB has a molecular mass $M_w =
291.44$~g~mol$^{-1}$ and a density of $\rho_{LC} =
0.996$~g~ml$^{-1}$. The 8CB, purchased from Frinton Lab, was
degassed under vacuum for about two hours in the isotropic phase
before used for pure and mixture samples. Spectroscopic grade
n-hexane (molecular mass of $86.18$~g~mol$^{-1}$, a density of
$0.6548$~g~ml$^{-1}$, and a boiling point of $342$~K) purchased
from EM Science was used without further purification. The 8CB and
n-hexane mixtures appear to be miscible up to an n-hexane mole
fraction of $\geq 0.12$. This was confirmed by polarizing
micrographs of the samples. Measurements were performed on samples
as a function of n-hexane mole fraction $x_{hex}$ ranging from $0$
(pure 8CB) to $0.12$.

High resolution ac-calorimetric measurements were carried out
using a homemade calorimeter. The calorimetric sample cell
consists of an aluminium envelop $15 \times 8 \times 0.5$~mm. To
prepare an envelop cell, a sheet of aluminum was cleaned using
successive application of water, ethanol, and acetone in an
ultrasonic bath and then was folded and sealed on three sides with
super-glue (cyanoacrylate). Once the cell was thoroughly dried,
the desired amount of liquid crystal followed by a relatively
large amount of n-hexane were introduced to the cell. The mass of
the sample and cell was monitored as the n-hexane was allowed to
evaporate slowly until the desired mass of the n-hexane was
achieved. At the point of the desired mass of the 8CB+hex mixture,
the envelop flap was quickly folded and sealed with the
super-glue.  When the filled cell was ready a $120$ $\Omega$
strain gauge and $1$ M$\Omega$ carbon-flake thermistor were
attached to opposite surfaces of the cell using GE varnish. The
cell was then mounted into the calorimeter, the details of which
can be found elsewhere \cite{Paul68, Dan97, Yao98}. In the
ac-mode, oscillating heating power $P_{ac}e^{i\omega t}$ is input
to the cell resulting in temperature oscillations with an
amplitude $ T_{ac}$ and a relative phase shift,
$\varphi=\Phi+\pi/2$, where $\Phi$ is the absolute phase shift
between $T_{ac}$ and the input power. Defining the heat capacity
amplitude as $C^\ast = P_{ac}/(\omega T_{ac})$, the specific heat
at a heating frequency $\omega$ can be expressed as
\begin{equation}
\label{eq:realC}
    C_{p}= \frac{C^{'}_{filled}-C_{empty}}{m_{s}} = \frac{C^{\ast}\cos(\varphi)-C_{empty}}{m_{s}}  \\
\end{equation}

\begin{equation}
\label{eq:Imc}
   C^{"}_{p} =  \frac{C^{"}_{filled}}{m_{s}} = \frac{C^{\ast}\sin(\varphi)-\frac{1}{\omega
   R_{e}}}{m_{s}}
    \end{equation}

where $C^{'}_{filled}$ and $C^{"}_{filled}$ are the real and
imaginary parts of the heat capacity, $C_{empty}$ is the heat
capacity of the empty cell, $m_{s}$ is the mass of the sample (in
the range of $15$ mg to $40$ mg), and $R_{e}$ is the external
thermal resistance between the cell and the bath.
Eq.~\eqref{eq:realC} and \eqref{eq:Imc} need small correction to
account the non-negligible internal thermal resistance as compared
to $R_{e}$ and this was applied to all samples \cite{Roshi04}. The
real part of the heat capacity indicates storage (capacitance) of
the energy whereas the imaginary part indicates the
loss(dispersion) of energy in the sample. Temperatures
corresponding to equilibrium, one-phase states exhibit a flat
imaginary part of heat capacity, i.e. $C^{"}_{p}=0$
\cite{Germano98}. Non-equilibrium dispersive regions, such as a
two-phase coexistence region where the latent heat is released,
have non-zero $C^{"}_{p}$.

\begin{figure}
\includegraphics[scale=1.0]{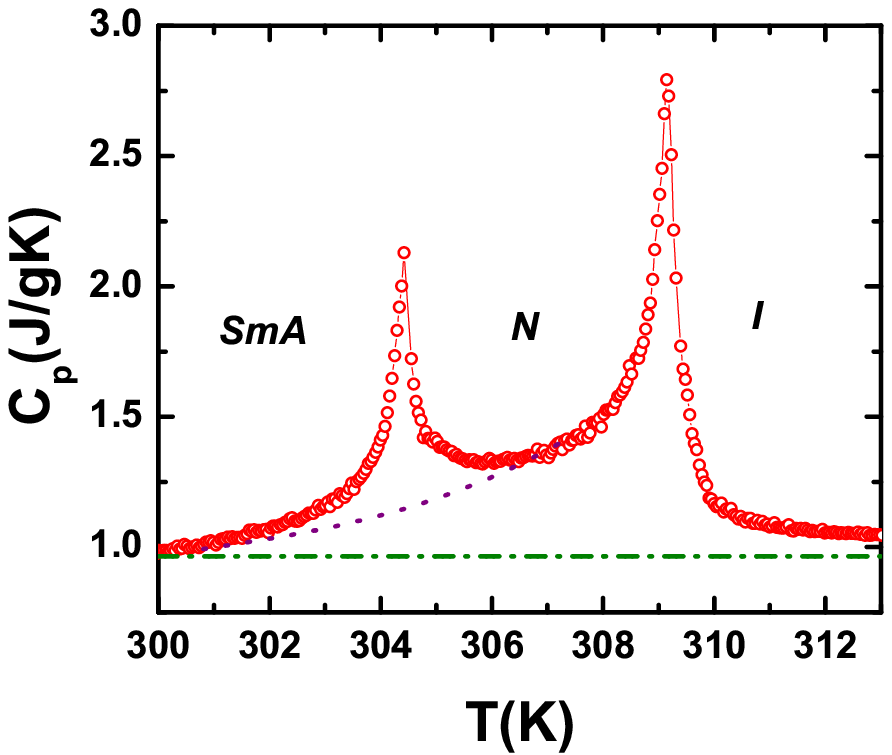}
\caption{ \label{Fig.1} Specific heat capacity for an 8CB+hex
sample on heating with $x_{hex}=0.02$. The dashed dotted line
represents the $C_{p}$ background, while the dashed curve acts as
$C^{wing}_{p}$ and represents the low temperature $I$-$N$ $C_{p}$
wing that would be expected in the absence of $N$-Sm$A$
transition. }
\end{figure}

Figure~\ref{Fig.1} illustrates the specific heat capacity
variation over an extended temperature range for the
$x_{hex}=0.02$ 8CB+hex sample. The dashed curve under the
$N$-Sm$A$ heat capacity peak represents the $I$-$N$ specific heat
capacity wing $C_{p}^{wing}$ expected in the absence of the
$N$-Sm$A$ transition. This wing is used to determine the excess
specific heat associated with the $N$-Sm$A$ phase transition

\begin{equation}
\label{eq:exheat}
 \delta C_{p} = C_{p}-C_{p}^{wing}.   \\
 \end{equation}

The  enthalpy change associated with a phase transition is defined
as
 \begin{equation}
 \label{eq:enthalpy}
 \delta H = \int\delta C_{p}  dT.\\
 \end{equation}

For a second-order or continuous phase transition, the limits of
integration are as wide as possible about the $\delta C_{p}$ peak
and gives the total enthalpy change ($\delta H$) associated with
the transition. But for a first-order transition the situation is
complicated due to the presence of a coexistence region  as well
as a latent heat $\Delta H$. The total enthalpy  change for a
weakly first order phase transitions  is the sum of the integrated
enthalpy and the latent heat, $\Delta H_{total} = \delta H +
\Delta H$. Due to partial phase conversion during a $T_{ac}$
cycle, typical $\delta C_{p}$ values obtained in the two-phase
coexistence region are artificially high and frequency dependent.
A simple integration of the observed $\delta C_{p}$ peak yields an
effective enthalpy change $\delta H^{\ast}$ for the first-order
transition which includes some of the latent heat contribution. If
we integrate the imaginary part of heat capacity given by
Eq.~\eqref{eq:Imc}, we can get the imaginary transition enthalpy
$\delta H^{"}$, which is the dispersion of energy in the sample
and is a proxy of latent heat associated with the transition. In
an ac-calorimetric technique the uncertainty in determining the
enthalpy is typically 10\% due to the uncertainty in the baseline
and background subtraction.


\section{RESULTS}
\label{sec:results}
\subsection{The $N$-Sm$A$ Heat Capacity}

\begin{figure}
\includegraphics[scale=0.75]{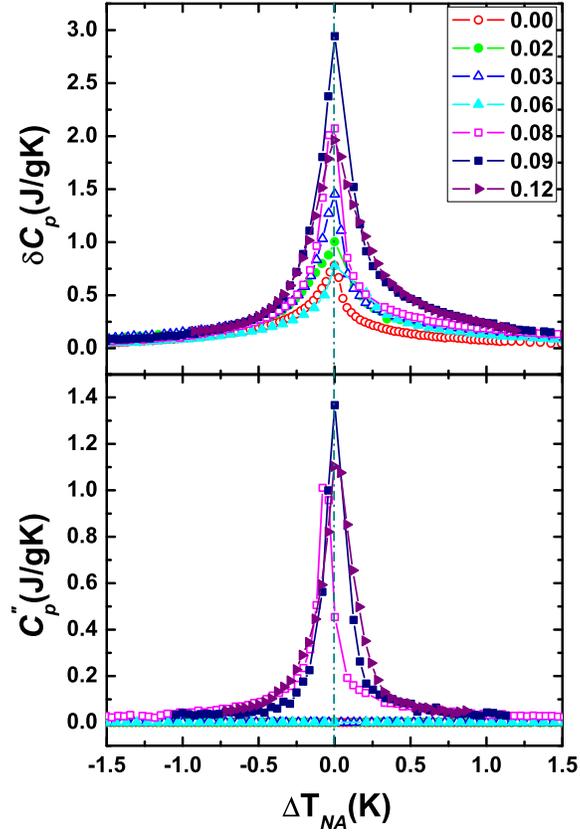}
\caption{ \label{Fig.2} Upper panel: The excess specific heat
$\delta C_{p}$ associated with the $N$-Sm$A$ transition on heating
as a function of temperature about $T_{NA}$ for pure and all
8CB+hex samples. See legend. Lower panel: The imaginary part of
heat capacity on heating for all samples as a function of
temperature about $T_{NA}$. }
\end{figure}

The resulting $\delta C_{p}$ data of the $N$-Sm$A$ transition on
heating for 8CB+hex and pure 8CB samples over a $\pm 1.5K$
temperature range window about the $\delta C_{p}$ peak is shown in
Fig.~\ref{Fig.2} (upper panel). As the mole fraction of n-hexane
increases, the $N$-Sm$A$ heat capacity peak becomes larger than
the pure $N$-Sm$A$ peak and with apparently larger wings on the
high temperature side of the peak. Figure~\ref{Fig.2} (lower
panel) shows the imaginary part of specific heat $C^{"}_{p}$ on
heating as a function of temperature about $T_{NA}$. For the
n-hexane mole fractions $x_{hex}\leq 0.06$, the $C^{"}_{p}$ is
flat, indicating the second-order nature of the transition. For
$x_{hex}\geq 0.08$, the $C^{"}_{p}$ reveals a peak indicating a
first-order behavior of the transition. As the mole fraction of
n-hexane increases beyond $x_{hex}\geq 0.08$, the $C^{"}_{p}$ peak
become broader with a two-phase co-existence region growing from
$\sim 0.35K$ at $x_{hex}=0.08$ to $\sim 0.85K$ at $x_{hex}=0.12$.

\begin{figure}
\includegraphics[scale=0.75]{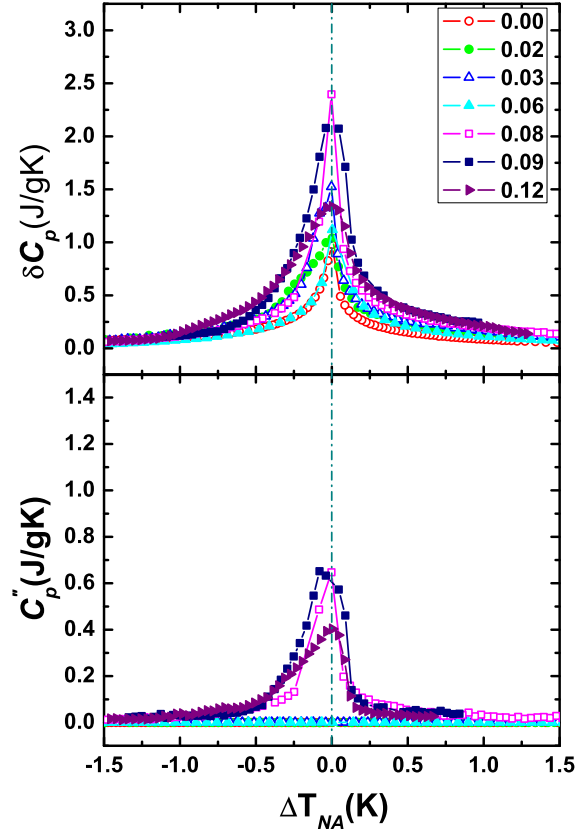}
\caption{ \label{Fig.3} Upper panel: The excess specific heat
$\delta C_{p}$ associated with the $N$-Sm$A$ transition on cooling
as a function of temperature about $T_{NA}$ for pure and all
8CB+hex samples. The definition of the symbols are given on the
inset. Lower panel: The imaginary part of heat capacity on cooling
for all samples as a function of temperature about $T_{NA}$.  }
\end{figure}

The $N$-Sm$A$ excess specific heat $\delta C_{p}$ (upper panel)
and imaginary part of heat capacity $C^{"}_{p}$ (lower panel)on
cooling are shown in Fig.~\ref{Fig.3}. On cooling, the $\delta
C_{p}$ peaks exhibit larger $C_{p}$ wings on both sides of
$T_{NA}$ but the low temperature wing appears progressively
smeared in temperature. In addition, the $\delta C_{p}$ on cooling
exhibits sharp peaks up to $x_{hex}=0.08$ then appears rounded for
$x > 0.08$. The $N$-Sm$A$ $C^{"}_{p}$ behavior on cooling is
similar to the heating scans in that $C^{"}_{p}=0$ through
$T_{NA}$ for $x_{hex} \leq 0.06$, then reveals a peak for
$x_{hex}\geq 0.08$. This indicates, as on heating, a cross-over
from continuous to first-order transition behavior. However, the
$C^{"}_{p}$ peaks for $x_{hex}\geq 0.08$ on cooling have markedly
different shape than on heating. Here, as the temperature
approaches $T_{NA}$ from above, a sharp jump preceded by a
relatively small wings occurs at $\sim 0.1K$ above $T_{NA}$ for
all 8CB+hex samples. As the temperature cools further, a long
$C^{"}_{p}$ tail is seen to a common trend at $\sim -0.25K$ for
$x_{hex}= 0.08$ and $\sim -0.4K$ for $x_{hex}= 0.09$ and $0.12$
below $T_{NA}$. The increase in the two-phase co-existence is
similar to that seen on heating.

\begin{figure}
\includegraphics[scale=0.75]{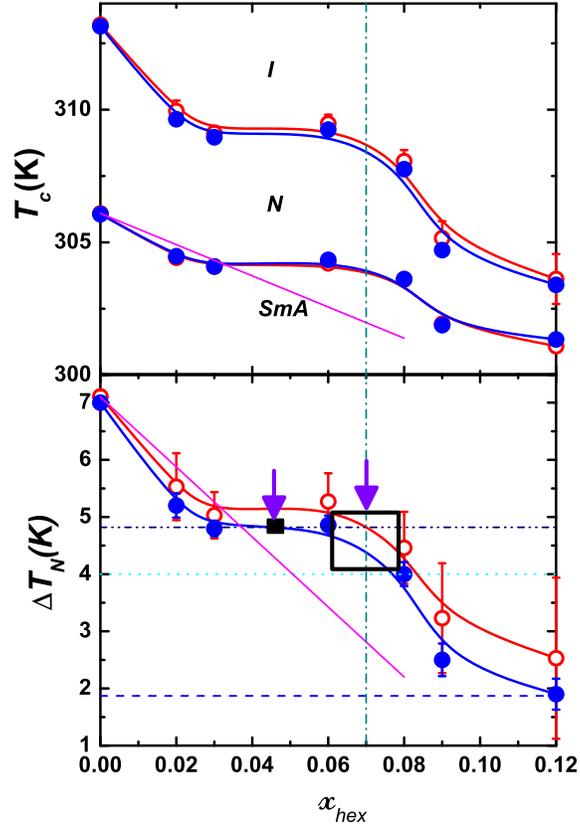}
\caption{ \label{Fig.4} Upper Panel: The $I$-$N$ and $N$-Sm$A$
phase transition temperatures on heating $(\circ)$ and cooling
$(\bullet)$ as a function of $x_{hex}$.  Solid lines are guides to
the eye. Lower panel: The nematic temperature range $\Delta T_{N}$
on heating $(\circ)$ and cooling $(\bullet)$ as a function of
$x_{hex}$. The closed and open rectangular boxes represent the
position of tricritical points for 8CB+chex \cite{Denolf07} and
8CB+hex systems respectively and the width and height of the boxes
represent the uncertainties on $x^{TCP}_{hex}$ and $\Delta
T^{TCP}_{N}$ respectively. The horizontal dashed, dashed dot, and
dot lines are nematic ranges for pure 9CB\cite{Qian98}, 8CB+chex,
and 8CB+10CB \cite{Lafouresse03} respectively at tricritical
point. The solid straight lines are transition temperature (upper
panel) and nematic range (lower panel) for 8CB+chex system
\cite{Denolf07}. }
\end{figure}

The $N$-Sm$A$ transition temperature $T_{NA}$ is defined as the
temperature of the $\delta C_{p}$ peak maximum and the $I$-$N$
transition temperature is taking at the lowest temperature of the
isotropic phase prior to entering the $I$+$N$ two-phase
coexistence region\cite{krishna1}. Figure~\ref{Fig.4} (upper
panel) shows the $I$-$N$ and $N$-Sm$A$ phase transition
temperatures as a function of $x_{hex}$. As $x_{hex}$ increases,
both transition temperatures decrease non-linearly with a bump at
$x_{hex} \sim 0.07$. Figure~\ref{Fig.4}(lower panel) shows the
nematic temperature range $\Delta T_{N} = T_{IN}-T_{NA}$ as a
function of $x_{hex}$ revealing a similar non-linear trend  with a
similar bump at the same $x_{hex}$. The horizontal dashed, dashed
dot, and dot lines represent nematic ranges for pure
9CB\cite{Qian98}, 8CB+chex\cite{Denolf07}, and 8CB+10CB
\cite{Lafouresse03}  at the tricritical point respectively. The
solid straight lines are the transition temperatures
(Fig.~\ref{Fig.4}-upper panel) and nematic range
(Fig.~\ref{Fig.4}-lower panel) for the 8CB+chex system
\cite{Denolf07}

Since continuous transition behavior is observed for $x_{hex}=
0.06$ and first-order behavior at $x_{hex}= 0.08$, a tricritical
point mole fraction is taken as $x^{TCP}_{hex}=0.07$ with the
corresponding nematic range at $\Delta T^{TCP}_{N}\simeq 4.63K$.
The vertical dashed line in both the panels of Fig.~\ref{Fig.4}
indicates $x^{TCP}_{hex}$ and a bold-bordered box in the
lower-panel gives the location of the cross-over point whose width
and height are the magnitude of uncertainties in $x^{TCP}_{hex}$
and $\Delta T^{TCP}_{N}$ respectively.

\begin{figure}
\includegraphics[scale=0.75]{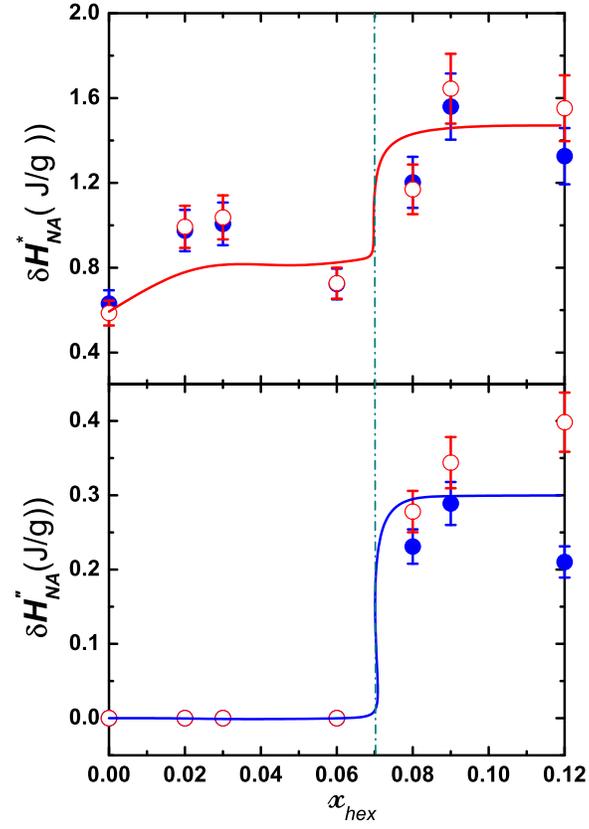}
\caption{ \label{Fig.5}  Upper panel: The total integrated $\delta
C_{p}$ ac-enthalpy $\delta H^{\ast}_{NA}$ on heating $(\circ)$ and
cooling $(\bullet)$ as the function of $x_{hex}$. Lower panel:
Integrated $C^{"}_{p}$ enthalpy $\delta H^{"}_{NA}$ on heating
$(\circ)$ and cooling $(\bullet)$ as the function of $x_{hex}$.
Solid lines are guides to the eye. }
\end{figure}

The effective $N$-Sm$A$ transition enthalpy $\delta H^{\ast}_{NA}$
was obtained by integrating $\delta C_{p}$ in the range $\pm3$K
about $T_{NA}$. The dispersive enthalpy, $\delta H^{"}_{NA}$ of
the $N$-Sm$A$ transition, available only for $x_{hex}\geq 0.08$,
and was obtained by integrating the $N$-Sm$A$ $C^{"}_{p}$ peak.
Since a fixed heating frequency was used, the non-zero $\delta
H^{"}_{NA}$ is only proportional to the transition latent heat.
The resulting $\delta H^{\ast}_{NA}$ and $\delta H^{"}_{NA}$ for
heating $(\circ)$ and cooling $(\bullet)$ scans as a function of
$x_{hex}$ for all 8CB+hex samples are shown in Fig.~\ref{Fig.5}.
The $\delta H^{\ast}_{NA}$ values show an overall increase in
value with increasing $x_{hex}$ and are consistent on heating and
cooling. A small apparent jump in $\delta H^{\ast}_{NA}$ is seen
at $\sim x^{TCP}_{hex}$. See Figure~\ref{Fig.5}(upper panel). The
$\delta H^{"}_{NA}$ exhibits a sudden jump from $0$ to $\sim
0.28$~$J/g$ at $x^{TCP}_{hex}$.

A summary of these results for 8CB+hex samples including pure 8CB
is tabulated in Table ~\ref{tab:summarytable}. Included are the
n-hexane molar fraction $x_{hex}$, the $N$-Sm$A$ transition
temperatures $T_{NA}$, nematic range $\Delta T_{N}$, integrated
enthalpy change $\delta H^{\ast}_{NA}$, imaginary enthalpy $\delta
H^{"}_{NA}$, McMillan ratio $MR$ and height of excess heat
capacity peaks $h_{M}$ for all the 8CB+hex samples including pure
8CB.

\subsection{Power-law Analysis of $N$-Sm$A$ phase transition}

Because the $\delta C_{p}$ for the $N$-Sm$A$ transition in 8CB+hex
remains continuous and sharp for $x_{hex}\leq x^{TCP}_{hex}$,  a
critical power-law analysis was performed. The usual power law
form in terms of reduced temperature, $|t|= |(T-T_{c})|/T_{c}$,
that is used to analyze the excess specific heat associated with
$N$-Sm$A$ transition is given by\cite{Garland94}
\begin{equation}
\label{eq:powerlaw}
 \delta C_{p}
=A^{\pm}|t|^{-\alpha}(1+D^{\pm}_{1}|t|^{\Delta_{1}})+B_{c},
\end{equation}

where $B_{c}$ is the critical background, $A^{\pm}$ are the
amplitudes above and below the transition, $D^{\pm}_{1}$ are the
correction-to-scaling amplitude with an exponent
$\Delta_{1}=0.524$ \cite{Garland94}. A full, non-linear, fitting
of Eq.\eqref{eq:powerlaw} to the $\delta C_{p}$ data was
attempted, but because the number of data close to the peak were
relatively sparse, these fits did not properly converge.

A simple power-law analysis procedure was employed in order to
estimate the variation of the critical exponent $\alpha$ as a
function of $x_{hex}$. This procedure begins by approximating
$T_{c}$ for each continuous $\delta C_{p}$ peak. This is done by
plotting a $log (\delta C_{p})$ vs $log(|t|)$ and choosing $T_{c}$
such that the high and low temperature wings appear linear and
parallel to each other for low $|t|$. The rounded and
non-power-law data points are easily determined and removed.

Figure~\ref{Fig.6} shows the resulting log-log plot of data above
and below $T_{c}$ for pure 8CB ($x_{hex}=0$) and the highest
concentration 8CB+hex sample that is continuous as determined by
$C^{"}_{p}$ ($x_{hex}= 0.06$). Now, a range of data up to
$|t_{max}|$ was chosen in order to perform a simple linear fit,
$log (\delta C_{p}) = log (A^{\pm}) - \alpha^{'}_{eff} log|t|$.
Here, $|t_{max}|$ varied smoothly from $8.9\times 10^{-4}$ for
pure 8CB to $1.8\times 10^{-3}$ for the $x_{hex} = 0.06$ 8CB+hex
sample. The resulting linear fits are shown in Fig.~\ref{Fig.6}
for data above and below $T_{c}$. The difference between $\alpha
^{'}_{eff}(T> T_{c})$ and $\alpha^{'}_{eff}(T<T_{c})$ is taken as
the uncertainty in $\alpha^{'}_{eff}$. The resulting
$\alpha^{'}_{eff}$ are not the true critical exponents because of
this simplified analysis. However, comparing the pure 8CB result
here to the literature value of $\alpha_{eff}=0.3$
\cite{Germano03}, a corrected $\alpha_{eff}$ for the 8CB+hex
samples is taken as an algebraic shift of $+0.17$, which is the
difference of $\alpha^{'}_{eff}- \alpha_{eff}$ for pure 8CB. This
procedure was applied for all samples from $x_{hex}= 0$ to $0.06$
and should reasonably approximate the $x_{hex}$ dependence of
$\alpha_{eff}$.

The resulting estimate of the $N$-Sm$A$ heat capacity effective
critical exponent as a function of $x_{hex}$ are shown in
Fig.~\ref{Fig.7}. Here, a linear rapid rise in $\alpha_{eff}$ is
seen as $x_{hex}$ increases from 0 to 0.04 then curving over for
$x_{hex} > 0.04$. The upward arrow in Fig.~\ref{Fig.7} is the best
estimate of $x^{TCP}_{hex}$ for this 8CB+hex system.


\section{DISCUSSION and CONCLUSION}
\label{sec:disc} The continuous $N$-Sm$A$ liquid crystal phase
transition has been studied using high-resolution ac-calorimetry
as a function of solvent dopant concentration. Multiple heating
and cooling cycles reproduce each other for $x_{hex}\leq 0.12$
along with no visual indication of phase separation support the
view that the 8CB+hex binary system remained mixed (n-hexane
miscible) for all samples studied here, without mechanical mixing.
This is supported also by x-ray studies of the smectic layer
spacing in 8CB+hex that showed phase separation for $x_{hex}> 1.0$
\cite{Rieker95}. In this work, the smectic layer spacing increased
with increasing $x_{hex}$ and was interpreted as a nano-scale
partitioning of n-hexane in between smectic layers.

A more recent calorimetric study of binary mixtures of 8CB with
various, low-molecular weight, solvents found dramatic changes to
the character  of the $N$-Sm$A$ phase transition \cite{Denolf07}.
In this work, the $N$-Sm$A$ transition approaches a tricritical
point linearly. However, this study used cyclo-hexane, that has
ring structure and employed continuous mixing during measurements
as a function of cyclohexane mole-fraction, $x_{chex}$. The
transition temperature $T_{NA}$ decreases linearly as $x_{chex}$
increases, the critical exponent $\alpha$ increases linearly from
0.31 (pure 8CB) to 0.50 at $x_{chex}= 0.046$, and the onset of a
$N$-Sm$A$ latent heat occurs smoothly at TCP, $x^{TCP}_{chex} =
0.046$. These results were modelled using mean-field
Landau-deGennes theory incorporating the nematic free-energy,
smectic free-energy , and a coupling between nematic and smectic
order parameters. This model was extended to account for the
solvent by adding a solvent mole-fraction coupling to $\psi^{2}$
and to $\psi^{2}\delta S$ to the total solvent free-energy.

Similar results were found in 8CB+biphenyl binary mixtures and a
Landau-de Gennes model that accounted for change in the LC elastic
constants with $x_{sol}$. However, TCP was not found in
8CB+biphenyl system \cite{DasGupta1}.

 In this present study, several important differences emerge. As
 $x_{hex}$ increases, $T_{NA}$ decreases as well as the nematic
 range $\Delta T_{N}$ in a non-linear way. The character of
 $N$-Sm$A$ transition remains continuous up to $x_{hex}\simeq 0.07$ where it
 appears to jump suddenly to a first-order transition. The bump in
 $T_{NA}$ and $\Delta T_{N}$ as well as the jump in $\delta
 H^{"}_{NA}$ all occur at $x^{TCP}_{hex}$.

\begin{figure}
\includegraphics[scale=0.9]{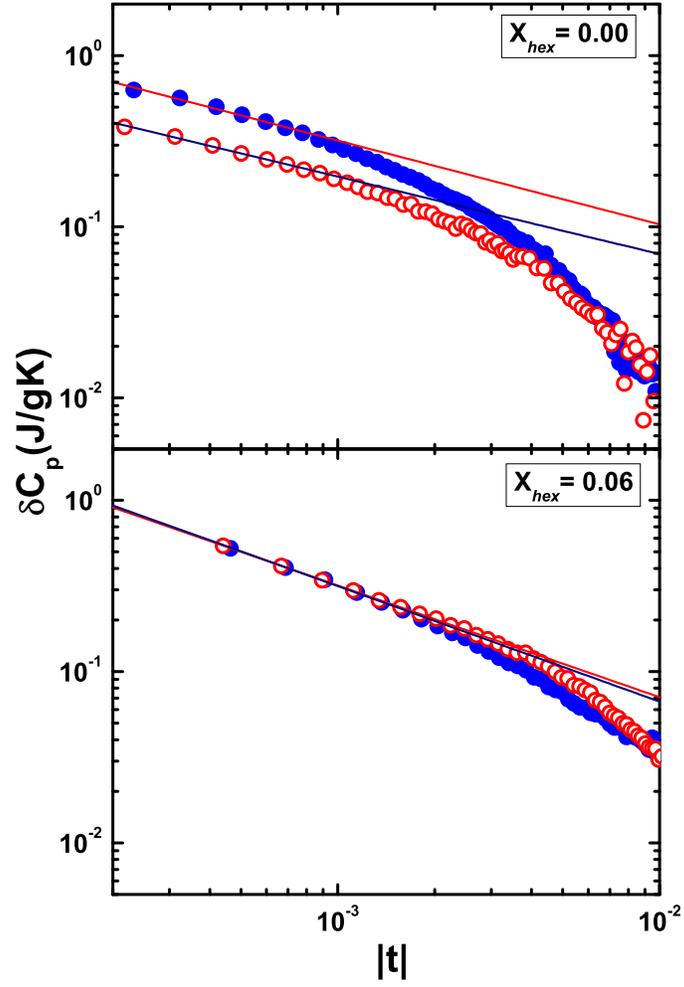}
\caption{ \label{Fig.6}  Upper panel: Excess specific heats
associated to $N$-Sm$A$ phase transition  as a function of reduced
temperature for pure 8CB for $T<T_{c}$ $(\bullet)$ and for  $T >
T_{c}$ $(\circ)$. Lower panel: Excess specific heats associated to
$N$-Sm$A$ phase transition  as a function of reduced temperature
for hexane mole fraction $x_{hex}=0.06$ for $T < T_{c}$
$(\bullet)$ and for  $T > T_{c}$ $(\circ)$. Slope of the straight
line in each graph gives the effective critical exponent
$\alpha_{eff}$.  }
\end{figure}

\begin{figure}
\includegraphics[scale=1.0]{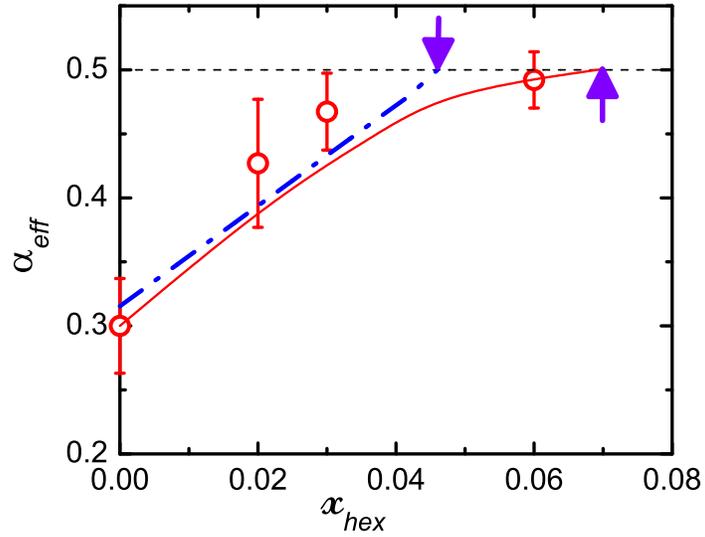}
\caption{ \label{Fig.7}   The effective critical exponent as a
function of hexane mole fraction. solid line represents the best
estimate of $\alpha_{eff}$ as a function of $x_{hex}$. This line
intersects the horizontal dashed line at the tricritical point,
$x_{hex}\simeq 0.07$. The dashed-dot line is $\alpha_{eff}$ for
8CB+chex system from reference \cite{Denolf07}. Vertical arrows
indicate the location of the tricritical points for 8CB+chex
(downward arrow) \cite{Denolf07} and 8CB+hex(upward arrow). }
\end{figure}

The critical behavior, estimated by the simple power-law analysis
presented here, evolves with $\alpha_{eff}$ initially increases
linearly as in the 8CB+chex system but then curves over to reach
$\alpha_{eff}=0.50$ at $x_{hex}\rightarrow x^{TCP}_{hex}$.
Qualitatively, the correction-to-scaling terms $D^{\pm}$ and the
amplitude ratio $A^{-}/A^{+}$ are changing their values towards
the tricritical values as a function of $x_{hex}$. Here, the
qualitative measurement of the amplitude ratio $A^{-}/A^{+}$ was
extracted examining the gap between two slope lines of the linear
fit of log-log plot of $\delta C_{p}$ vs $|t|$ (Fig.~\ref{Fig.6})
and the curvature of the curve at high $|t|$ was observed to get
qualitative measure of $D^{\pm}$.

The addition of n-hexane in 8CB creates the random dilution effect
which causes the decrease in transition temperature $T_{NA}$ and
nematic range $\Delta T_{NA}$. The experiment was done without
stirring the sample which may cause some phase separation in
microscopic or even in nanoscopic scale. These dilution and
microphase separation effects, may cause to develop the two
competing interactions which cause the non-linearity in the
transition temperature $T_{NA}$, nematic range $\Delta T_{N}$,
effective critical exponent $\alpha_{eff}$ and jump in the
imaginary enthalpy $\delta H^{"}_{NA}$. These effects also cause
the change in coupling between the order parameters $\psi$ and $Q$
which consequently change the order of the $N$-Sm$A$ phase
transition from continuous to first order with a critical point at
$x^{TCP}_{hex} \simeq 0.07$.

We have undertaken a detailed calorimetric studies on the effect
of non- mesogenic, low molecular weight solvent(hexane) on
octylcyanobiphenyl(8CB) phase transitions with emphasis on the
most extensively studied but controversial $N$-Sm$A$ phase
transition.  The addition of the hexane on 8CB dilutes the mixture
and changes the intermolecular potential. The microscopic phase
separation and dilution effect due to the introduction of n-hexane
in 8CB cause the change in magnitude of $S$-$\psi$ and
$\psi-\delta\hat{n}$ couplings which consequently change the phase
transition behavior. The result obtained reveals new aspect of the
effect of non-mesogenic disorder on the liquid crystal transition.


\begin{acknowledgements}
This work was supported by the Department of Physics at WPI.
\end{acknowledgements}


\newpage


\begin{table*}
\caption{ \label{tab:summarytable} Summary of the calorimetric
results for pure and all 8CB+hex samples on heating. Shown are
hexane molar fraction $x_{hex}$, $N$-Sm$A$ transition temperature
$T_{NA}$, nematic range $\Delta T_{N}=T_{IN}-T_{NA}$( in Kelvin),
integrated enthalpy change $\delta H^{\ast}_{NA}$, imaginary
enthalpy $\delta H^{"}_{NA}$ (in J/g),  McMillan's Ratio $MR$, and
heat capacity maximum $h_{M} \approxeq  \delta C^{max}_{p}(N-A)$
in $(JK^{-1}g^{-1})$. }

\begin{ruledtabular}
\begin{tabular}{@{\extracolsep{20 pt}} ccccccc}
 $x_{hex}$ & $T_{NA}$ & $\Delta T_{N}$  & $\delta H^{\ast}_{NA}$ & $\delta H^{"}_{NA}$ & $MR$  & $h_{M}$ \\
\hline
 $0.00$   &  $306.09\pm0.06$ &  $7.11\pm0.11$   & $0.59\pm0.06$  &  $-$           &  $0.977$   & $ 0.78$  \\
 $0.02$   & $304.42\pm0.09$  & $5.53\pm0.59$    & $0.99\pm0.10$  &  $-$           &   $0.982$  & $1.00$ \\
 $ 0.03$  & $304.09\pm0.08$  & $5.03\pm0.41$    & $1.04\pm0.10$  &  $-$           &  $0.984$   & $1.45$ \\
 $ 0.06$  & $304.21\pm0.14$  & $5.27\pm0.50$    & $0.73\pm0.07$  &  $-$           &  $0.983$   & $0.78$ \\
 $0.08$   &  $303.60\pm0.16$ &  $4.46\pm0.63$   & $1.17\pm0.12 $ &  $0.28\pm0.03$ &  $0.986$   & $2.07$ \\
 $0.09$   &  $301.93\pm0.17$ &  $3.23\pm0.96$   & $1.64\pm0.16$  & $0.29\pm0.03$  & $0.989$    & $2.94$  \\
 $0.12$   &  $301.09\pm0.08$ & $2.53\pm1.41$    & $1.55\pm0.15$  & $0.40\pm0.04$  & $0.989$    & $1.96$  \\
\end{tabular}
\end{ruledtabular}

\end{table*}

\end{document}